\documentclass{an}
\usepackage{times}
\usepackage{amssymb}
\usepackage{graphicx}
\begin{document}
\newcommand{\beq}{\begin{equation}}
\newcommand{\eeq}{\end{equation}}
\newcommand{\bea}{\begin{eqnarray}}
\newcommand{\eea}{\end{eqnarray}}
\newcommand{\bfig}{\begin{figure}[!t]}
\newcommand{\efig}{\end{figure}}
\newcommand{\bfigc}{\begin{figure}[!t]\begin{center}}
\newcommand{\efigc}{\end{center}\end{figure}}
\newcommand{\bt}{\begin{table}}
\newcommand{\et}{\end{table}}
\newcommand{\btu}{\begin{tabular}}
\newcommand{\etu}{\end{tabular}}
\newcommand{\bc}{\begin{center}}
\newcommand{\ec}{\end{center}}
\newcommand{\bi}{\begin{itemize}}
\newcommand{\ei}{\end{itemize}}
\newcommand{\bd}{\begin{description}}
\newcommand{\ed}{\end{description}}
\def\Rad{\tilde\omega}
\def\sol{ \ \mathrm{M}_\odot}
\newcommand{\tensor}[1]{\buildrel\leftrightarrow\over #1}
%
\Pagespan{1024}{1027}
\Yearpublication{2006}
\Yearsubmission{2006}
\Month{10}
\Volume{327}
\Issue{10}
\DOI{10.1002/asna.200610685}

\title{The onset of General Relativity: Gravitationally redshifted emission lines}
\author{Andreas M\"uller\thanks{Corresponding author:
  amueller@mpe.mpg.de\newline}}
\titlerunning{Gravitationally redshifted emission lines}
\authorrunning{A. M\"uller}
\institute{Max--Planck--Institut f\"ur extraterrestrische Physik, PO box 1312, D--85741 Garching, Germany}
\received{2006, Sep 1} 
\accepted{2006 Oct 4}
\publonline{2006 Nov 14}
%
%
%
%
\abstract{We study and quantify gravitational redshift by means of relativistic ray tracing simulations 
of emission lines. The emitter model is based on thin, Keplerian rotating rings in the equatorial plane 
of a rotating black hole. Emission lines are characterised by a generalized fully relativistic Doppler 
factor or redshift associated with the line core.
Two modes of gravitational redshift, shift and distortion, become stronger with the emitting region 
closer to the Kerr black hole. Shifts of the line cores reveal an effect at levels of 0.0015 to 60~\% at 
gravitational radii ranging from $10^{5}$ to 2. The corresponding Doppler factors range from 0.999985 to 
0.4048. Line shape distortion by strong gravity, i.e.\ very skewed and asymmetric lines occur at radii 
smaller than roughly ten gravitational radii.
Gravitational redshift decreases with distance to the black hole but remains finite due to the 
asymptotical flatness of black hole space--time. The onset of gravitational redshift can be 
tested observationally with sufficient spectral resolution. Assuming a resolving power of $\sim100000$, 
yielding a resolution of $\approx0.1$ {\AA} for optical and near--infrared broad emission lines, the 
gravitational redshift can be probed out to approximately 75000 gravitational radii.
In general, gravitational redshift is an indicator of black hole mass and spin as well as for 
the inclination angle of the emitter, e.g.\ an accretion disk. We suggest to do multi--wavelength observations 
because all redshifted features should point towards the same central mass.}
\maketitle
%
%
%
\section{Introduction}
General Relativity (GR) teaches us that electromagnetic radiation is influenced by masses in both, 
photon energy and spectral flux. This gravitational redshift effect is remarkably strong for 
compact masses such as black holes and responsible for their blackness. It is well established 
that black hole candidates in the universe hide in X-ray binaries (XRBs; Bolton 1972; van der Klis 
1994, 2000) and active galactic nuclei (AGN; Lynden--Bell 1969, 1971; Rees 1984; Netzer 2003). The 
existence of intermediate--mass black holes (IMBHs) in globular clusters (Baumgardt et al. 2003; 
Gebhardt, Rich \& Ho 2005) and ultra--luminous X--ray sources (Fabbiano 1989; King et al. 2001;
Pakull, Gris\'e \& Motch 2005; Miller 2005) is still under debate. Of course, gravitational 
redshift plays a major role in all these black hole mass families.
According to the AGN standard model, there exists the so--called broad--line region (BLR) that 
is located between $\sim 10^{3}$ to $\sim 10^{5} \ r_\mathrm{g}$ 
away from the center. Here the gravitational radius is defined as $r_{\rm g}= {\rm G}M/{\rm c}^{2}$ 
with Newton's constant ${\rm G}$, vacuum speed of light ${\rm c}$ and black hole mass $M$ (we set 
${\rm G}={\rm c}=1$ throughout the paper). Doppler broadened optical emission lines with widths of 
typically 10$^{3}$--10$^4$ km\,s$^{-1}$ form in the BLR (Woltjer 1959). It has been first speculated 
by Netzer (1977) that gravitational redshift may influence optical lines causing line asymmetries. 
Meanwhile, this has been reported in a number of sources, e.g.\ Peterson et al. (1985), 
Zheng \& Sulentic (1990), and Kollatschny (2003). The gravitationally redshifted iron K fluorescence 
lines are very prominent X-ray features and have been reported for AGN and XRBs 
(Fabian et al. 1989; Tanaka et al. 1995; Martocchia et al. 2002; Miller et al. 2004). First calculations
of such lines around a maximally rotating black hole were performed by Laor (1991). A recent study by 
Nandra (2006) supports the idea that broad relativistic iron K lines are blended by non--relativistic 
narrow lines originating most probably from the cold dust torus.
The Schwarzschild factor allows for immediately estimating gravitational redshift for static emitters. 
However, it is poorly understood how this effect evolves with distance to the black hole if a reasonable 
emitter kinematics is assumed. In this paper, we study the gravitational redshift over a large range of 
distances from the central black hole. We quantify the relativistic gravitational redshift on emission 
lines until GR fades beyond the current observable limit. The investigation is carried out in a very general 
form by discussing the observed line profile as a function of the generalized GR Doppler factor ($g$--factor) 
for Kerr black holes. Details of this work are presented in M\"uller \& Wold (2006). Previous work on the 
that topic was done by Cunningham (1975) and Corbin (1997).
\section{Method and model}
We compute emission lines with the \textit{Kerr Black Hole Ray Tracer} (KBHRT) that maps emitting 
points in the equatorial plane of the black hole to points on the observer's screen. In a further 
step, we calculate the spectral line fluxes by numerical integration. All relativistic effects such 
as gravitational redshift, beaming and lensing are included, but higher order images are not 
considered. The solver has been presented in M\"uller \& Camenzind (2004, MC04 hereafter).

We analyse the simulated line profiles by considering the $g$-factor which generally reads
\beq \label{eq:g}
g=\nu_\mathrm{obs}/\nu_\mathrm{em}=\lambda_\mathrm{em}/\lambda_\mathrm{obs}=\frac{1}{1+z},
\eeq
where $\nu$ and $\lambda$ denote frequency and wavelength, respectively, and the redshift is $z$. 
Emitter's and observer's frame of reference are indicated by subscripts 'em' and 'obs'. The
value of $g$ signals redshift ($g<1$), blueshift ($g>1$), and no net shift ($g=1$). 
We characterize simulated line profiles by the line core energy (or line centroid)
\beq \label{eq:gcore}
g_\mathrm{core}=\frac{\sum_i \ g_\mathrm{i}\,F_\mathrm{i}}{\sum_i \ F_\mathrm{i}},
\eeq
i.e.\ the $g$--factor that belongs to the flux weight of the whole line. This is a 
suitable quantity to quantify redshifts and to investigate the transition from 
Einsteinian to Newtonian gravity .
Gravitational redshift in the weak field regime establishes pure shifts of spectral 
features without changing their intrinsical shape. However, gravitational redshift
in the strong field regime -- \textit{strong gravity} -- produces remarkable distortions
of spectral shapes if the rest frame feature is compared to its analogue in the observer's 
frame. Distortion is a key feature of relativistic spectra exhibiting very skewed and 
asymmetric line profiles (Fabian et al. 1989; Popovic et al. 1995; Tanaka et al. 1995). 
These effects are important only very close to the black hole. 
The emitter model consists of stationary Keplerian rotating and geometrically thin rings that
lie in the black hole's equatorial plane. In the course of studying gravitational redshift it is 
better to avoid blueshift effects. Hence, a face--on orientation of the emitting region, 
$i=0^\circ$, is a reasonable choice. In that way, the logitudinal Doppler effect is suppressed 
because there is no relative emitter motion along the line of sight. However, we still 
have to take into account the transverse Doppler effect in this case. This second order GR effect 
comes into play for relativistic speeds of the emitter. We set an inclination angle of $i=1^\circ$ 
for numerical reasons. Further, other parameters are chosen to be black hole spin at Thorne's 
limit $a=0.998\,M$ (Thorne 1974) and a prograde ($+$) Keplerian velocity field of emitters, 
$\Omega^+_\mathrm{K}=\sqrt{M}/(\sqrt{r^3}+a\sqrt{M})$.
The emissivity for the rings is modelled by a Gaussian radial emissivity profile that is endowed 
with an emission peaking at $R_\mathrm{peak}$. It is 
$\epsilon(r)\propto\exp\left(-(r-R_\mathrm{peak})^2/\sigma_\mathrm{r}^2\right)$ as introduced by 
MC04. The parameter $\sigma_\mathrm{r}$ controls the width of the Gaussian or 
the size of the emitting region and is chosen to be 0.2 for all runs. A localized Gaussian 
emissivity set in this way mimics a thin and narrow luminous annulus with $\simeq 1 \ \mathrm{r_g}$ 
distance between inner and outer edge. 
In our procedure the rings are shifted from large to small distances to the black hole. We assume 
that the radial range of interest is $2 \ \mathrm{r_g} \leq R_\mathrm{peak} \leq 100000 \ \mathrm{r_g}$. 
For each simulated emitting ring we determine the line core redshift $z_\mathrm{core}$ by using
Eqs. (\ref{eq:gcore}) and (\ref{eq:g}). If we assign $R_\mathrm{peak}$ to each value of 
$z_\mathrm{core}$, we are able to quantify the gravitational redshift as a function of distance 
to the rotating black hole, i.e.\ the redshift gradient. 
\section{Two modes of gravitational redshift: shift and distortion} \label{sec:twomod}
%
%
\bfigc
	\rotatebox{0}{\includegraphics[width=0.5\textwidth]{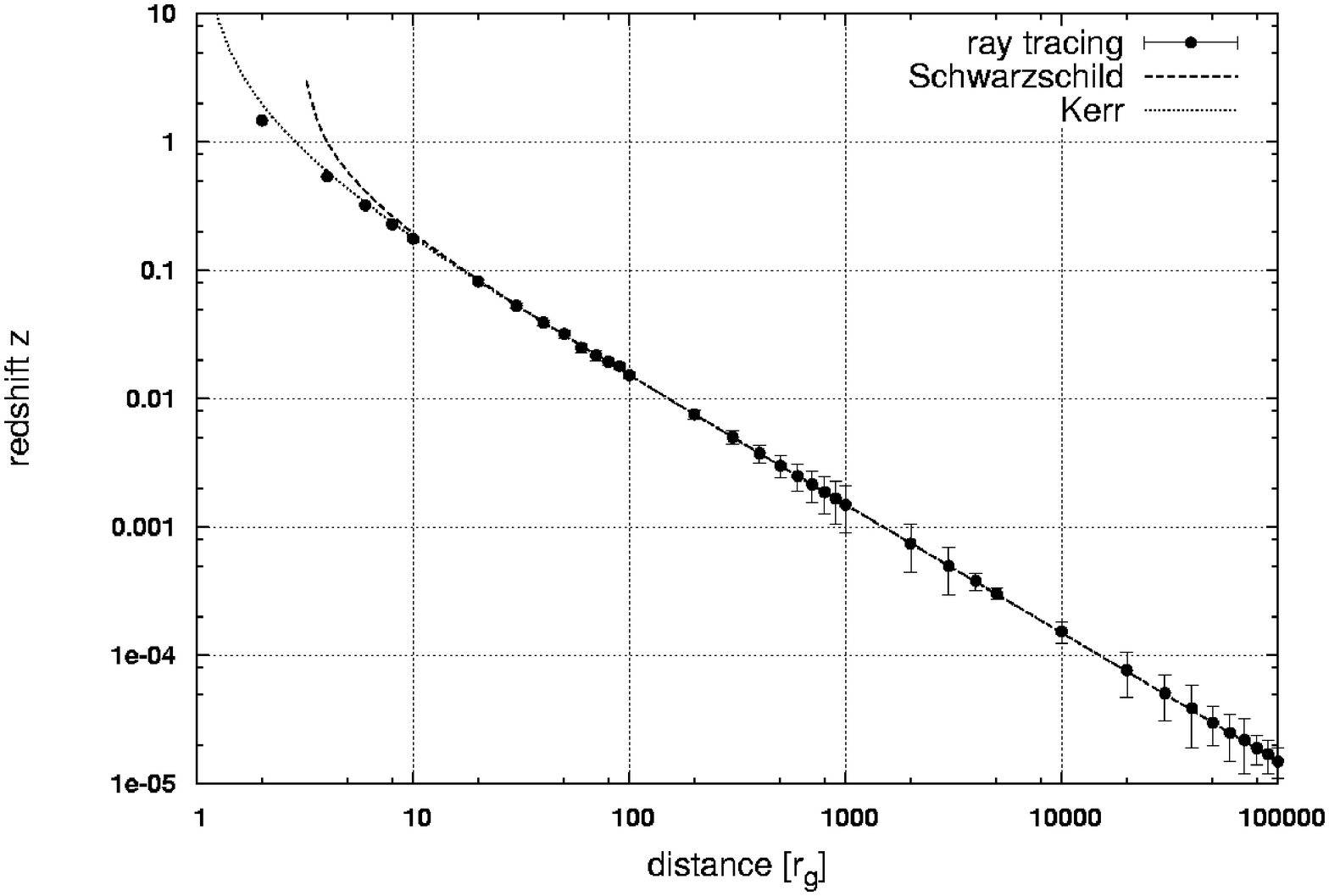}}
	\caption{Decay of gravitational redshift $z$ with increasing distance to the rotating 
	black hole with $a=0.998\,M$. The ray tracing simulations (\textit{filled circles}) are 
	compared to the analytical redshifts in Schwarzschild and the Kerr geometry, respectively
	(see text for analytical expressions). The inclination angles of the rings amount to 
	$1^\circ$. The error bars reflect that ray tracing delivers slightly varying results 
	depending on numerical resolutions for emitting area and/or spectrum.} \label{fig:z-r-plot}
\efigc
%
Fig. \ref{fig:z-r-plot} displays the core redshifts of emission lines, $z_\mathrm{core}$, as a 
function of $R_\mathrm{peak}$ for rings inclined to $i=1^\circ$. The redshifted lines are dominated 
by gravitational redshift by construction. The curve shows that the redshift approaches 
$z\rightarrow 0$, i.e.\ that $g\rightarrow 1$, at distances of a few thousand gravitational radii 
from the black hole. This is the regime of nearly flat space--time and Newtonian physics. But 
approaching the black hole, space--time curvature becomes more significant: $z$ grows rapidly and 
$g$ approaches zero.
We compare the simulation results with analytical expressions for the redshift $z$. These redshifts 
follow from Eq. (\ref{eq:g}) and the fact that $g_{i=0^\circ}=\alpha/\Gamma$ for a face--on situation, 
$i=0^\circ$ (c.f. Lind \& Blandford 1985).
Here $\alpha$ denotes the lapse function of the Kerr geometry and $\Gamma$ is the Lorentz factor. The 
redshift dependence on $\Gamma$ accounts for the transverse Doppler effect that is important even for 
rings tilted to face--on orientation. The lapse function $\alpha$ in Boyer--Lindquist form holds (if 
restricted to the equatorial plane, $\theta=\pi/2$; see MC04, Eq. 3):
\beq
\alpha=\sqrt{\frac{r(r^2-2Mr+a^2)}{r^3+a^2(r+2M)}}.
\eeq
This function reduces to the famous Schwarzschild factor $\alpha=\sqrt{1-2M/r}$ for $a=0$. The emitter 
has a simple velocity field represented by a Keplerian rotation so that the Lorentz factor reads
\beq
\Gamma=\left(1-v^{(\Phi)}\,v^{(\Phi)}\right)^{-1/2}.
\eeq
The orbital velocity is expressed in the locally non-rotating frame (Bardeen 1970a, 1970b) by 
\beq
v^{(\Phi)}=\frac{\tilde{\omega}}{\alpha}\left(\Omega^+_{\rm K}-\omega\right),
\eeq
with cylindrical radius $\tilde{\omega}$ and frame--dragging frequency $\omega$. Finally, we 
therefore arrive at redshift $z_{i=0^\circ}(r)=(\Gamma-\alpha)/\alpha$ that can be expanded 
into a binomial for $r\gg GM/c^2$. Then, we get $z\propto\,r^{-1}$ in the far field. This is 
true for both, Kerr and Schwarzschild black holes. The resulting redshift curve for a 
non--rotating black hole with $a=0$ (\textit{dashed curve}) versus a rotating black hole with 
$a=0.998\,M$ (\textit{dotted curve}) in Fig. \ref{fig:z-r-plot} show that deviations due to 
frame--dragging become necessary only very close to the black hole\footnote{In this Keplerian 
model plausible redshifts are only computed for radii greater than the marginally stable orbit, 
$r_{\rm ms}(a)$}. We return to this point in the next section. The ray 
tracing simulations confirm the expectations based on simple analytics: The gravitational redshift
decays with increasing distance. But all classical black hole solutions are asymptotically flat. Hence, 
there is no finite distance at which the gravitational redshift vanishes exactly. However, observability 
poses a limit to what is practical since the resolving power of a spectrograph constrains the amount of 
gravitational redshift that can be detected. Assuming a spectral resolution of 0.1 \AA\, 
for H$\beta$ as is obtainable, e.g.\ by VLT instruments, the corresponding critical value of the 
$g$-factor is $g=0.999979$. In our model this shift occurs at a radius of $\sim 75000 \ r_\mathrm{g}$. 
For supermassive black holes (SMBHs) of 10$^{7}$--10$^{8}$ M$_{\odot}$ this radius corresponds to 0.05--0.5 pc, 
whereas for stellar--mass black holes of $\sim 10$ M$_{\odot}$ it is 0.01~AU. Though, an observational 
complication is that at large distances to the black hole the line shifts are not dominated by 
gravitational redshift.
Gravitational redshift not only occurs as a pure shift: relativistic emission lines are also
deformed in intrinsic shape. But this mode only occurs in sufficient proximity of the emitter 
to the black hole. Hence, we refer to this as strong gravity. It is suitable to change the
inclinations to higher angles to demonstrate this second effect. Therefore, we set $i=75^\circ$ 
(any other simulation parameters remain) and investigate the characteristic broad line profile 
with the two relic Doppler peaks for various distances of the rings to the black hole. We make 
use of two quantities to illustrate the distortion effect. The first quantity termed 
$g_\mathrm{rp}$ is the $g$--factor associated with the red relic Doppler peak that is shifted 
to lower energies as the ring approaches the hole, displayed in Fig. \ref{fig:g_rp-i75}. This 
plot can be used to read the half--energy radius (i.e.\ the radius where the $g$--factor 
associated with the red Doppler peak is exactly $1/2$) associated with the red relic Doppler 
peak. At $i=75^\circ$ this quantity can be found at $\simeq 5 \ \mathrm{r_g}$ and documents 
that strong gravity is important only very close to the black hole. 
%
\bfigc
	\rotatebox{0}{\includegraphics[width=0.5\textwidth]{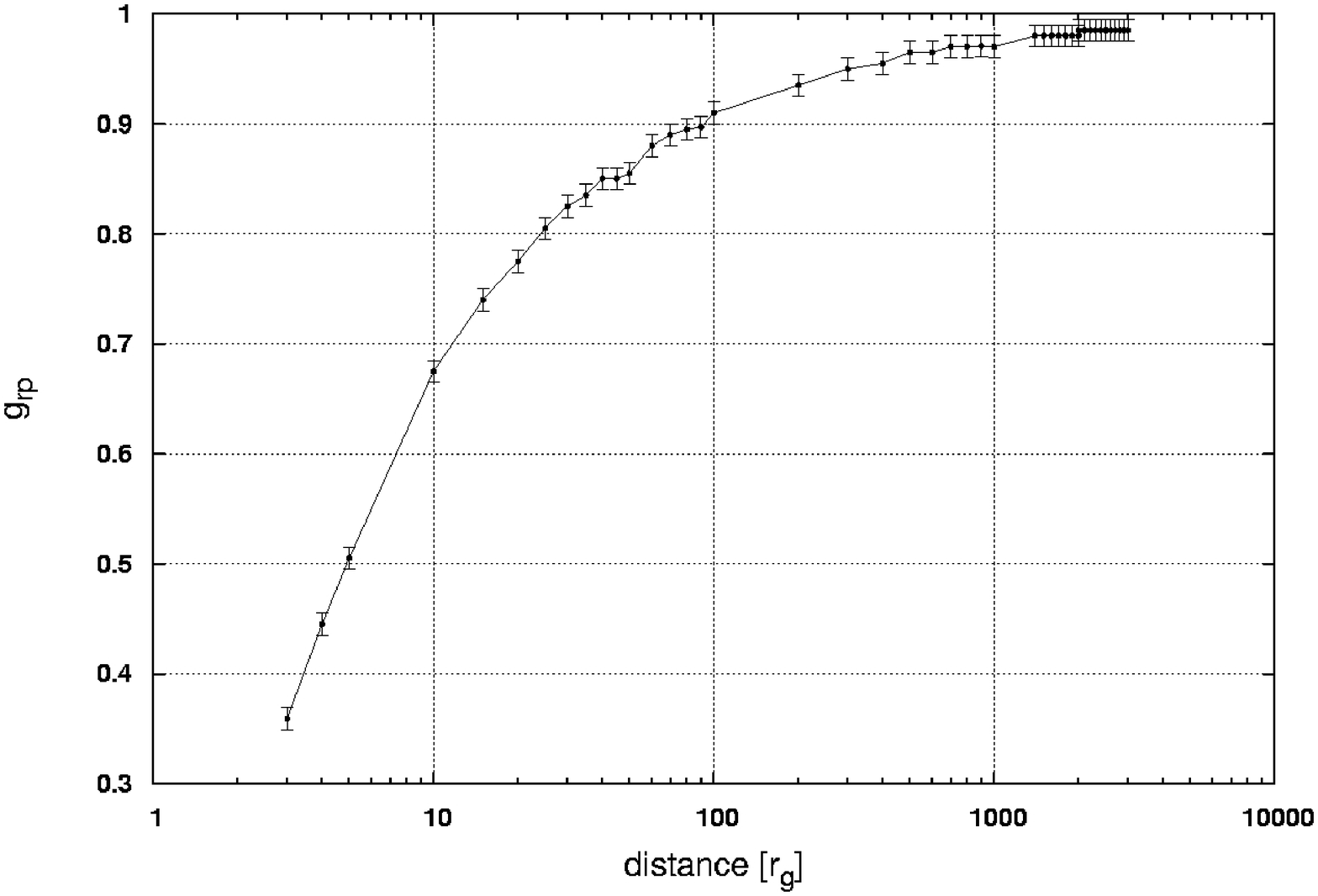}}
 	\caption{Distortion of the red relic Doppler peak for rings satisfying $i=75^\circ$: With 
	decreasing radius the red peak can be found at lower peak energies due to strong gravitational 
	redshift. Very close to the black hole -- $r\lesssim 3 \ \mathrm{r_g}$ at this specific 
	inclination -- these distortions are such strong that the red Doppler peak is highly blurred 
	and vanishes effectively in the line profile.} \label{fig:g_rp-i75} 
\efigc
%
\bfigc
	\rotatebox{0}{\includegraphics[width=0.5\textwidth]{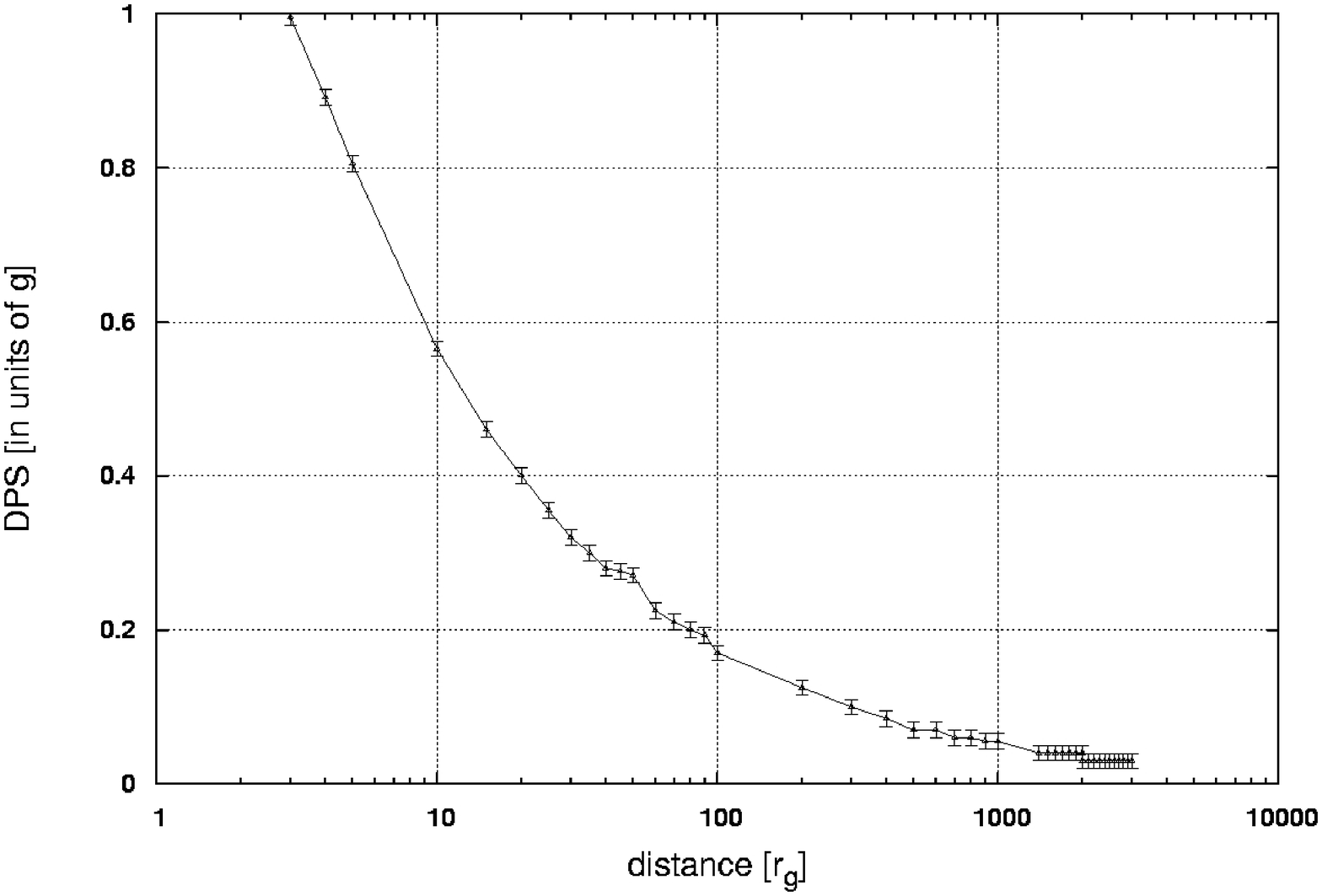}}
	\caption{Doppler peak spacing (DPS) in units of $g$ derived from line emitting rings at 
	$i=75^\circ$. Far away from the black hole both peaks approach significantly and the 
	double--peaked line profile becomes very narrow: At $\sim 3000 \ \mathrm{r_g}$ the peak 
	difference in $g$ only amounts to $\sim 0.03$. } \label{fig:DPS-i75}
\efigc
The distortion of the intrinsic line profile by gravitational redshift can be seen from $g_\mathrm{rp}$ 
alone or from the second quantity called Doppler peak spacing (DPS) that is computed from the g--factors 
associated with the red peak (rp) and the blue peak (bp): ${\rm DPS}=g_\mathrm{bp}-g_\mathrm{rp}$. 
Fig. \ref{fig:DPS-i75} shows that the spacing between the Doppler peaks does not remain constant as the 
emitter approaches the black hole: The DPS value rises quickly, i.e.\ the line is stretched. Gravitational 
redshift causes an additional suppression in flux so that close to the black hole the emission line from 
an intermediately to highly inclined ring is asymmetric and skewed.
\section{Effects of black hole rotation} \label{sec:rot}
A \textit{rotating} black hole gains support from black hole growth theories: Shapiro (2005) 
reviewed spin values of SMBHs in the local universe to be at either 
$a\simeq 0.95\, M$ (MHD disk) or even $a\simeq M$ (standard thin gas disk). But rotation of 
space--time decreases very steeply with distance to the black hole,\ $\omega\propto r^{-3}$. 
Fig. \ref{fig:z-r-plot} already documents a difference between the Kerr and the Schwarzschild 
redshifts only for, $r\lesssim 10 \ \mathrm{r_g}$. Therefore, black hole rotation can only be 
probed with spectral features originating in regions very close to the black hole, like X--ray 
iron K lines. In contrast, optical BLR emission lines in AGN are not suited for probing black 
hole rotation in that manner. These optical and X--ray lines together cover a radial scale of 
five orders of magnitude. Only larger multi--wavelength data sets for each particular source allow 
for testing gravitationally redshifted lines in different spectral ranges. All these data should 
point towards the same central mass that could be determined this way. High--quality data from 
the black hole's vicinity could even give rise for its spin.
Kollatschny (2003) and also M\"uller \& Wold (2006) demonstrated for the Narrow--line Seyfert--1 
galaxy Mrk 110 that gravitational redshift studies also reveal the inclination angle of the inner 
accretion disk if the flat, Keplerian BLR scenario holds. This offers a new powerful black hole
mass estimator besides reverberation mapping, $M$--$\sigma$ relation, stellar velocity dispersion 
or Maser studies. 
\section{Conclusions} \label{sec:conc}
We analyse emission line cores that originate from Keplerian rotating rings at 2 to 100000 $r_\mathrm{g}$ 
distance to a rapidly spinning black hole by using relativistic Kerr ray tracing simulations. These 
simulations confirm the expectation that the redshift gradient follows a simple 
$z\approx\frac{1}{\mathrm{distance[r_g]}}$ law as a straightforward consequence of the Schwarzschild factor. 
However, closer to the black hole the behaviour is not that simple and redshift is influenced by non-linear 
effects and black hole spin.
Generally, gravitational redshift occurs in two modes: One regime starts at larger distances from the black 
hole and shifts only the line as a total feature while conserving its intrinsic shape. The other regime, which 
is the strong gravity regime, dominates at $r\lesssim 10 \ \mathrm{r_g}$. Emission lines that form in the 
vicinity of the black hole are strongly deformed and suppressed. 
GR predicts that gravitational redshift is a ubiquitous phenomenon. However, the underlying redshift may be 
blurred by blueshift effects, e.g.\ from outflows. If the amount of blueshift is known or negligible, the 
verification of gravitational redshift is only detector--dependent. Assuming currently feasible optical 
resolving powers, we find a limiting distance of $75000 \ \mathrm{r_g}$ where gravitationally redshifted 
optical lines can be probed. But competing effects and more complex physics are likely to be involved so that 
the verification remains challenging.
We suggest to intensify high--resolution multi--wavelength observations in varying distance around black 
hole candidates so that the central mass can be derived from systematic spectral shifts. If our flat 
Keplerian model applies for line emitting regions, it is even possible to determine the inner inclination 
angle. Black hole spin can only be probed by spectral features that form sufficiently close to the center, 
e.g.\ by iron K$\alpha$ lines.
Our study holds for any black hole of arbitrary mass, i.e.\ stellar--mass black holes in XRBs, IMBHs that 
may be found in ultra--luminous X--ray sources or globular clusters, and SMBHs in the cores of galaxies.
%
%
%
\begin{acknowledgements}
I wish to thank an anonymous referee who helped to improve this paper. I am grateful to the organizers 
and participants of the ESAC/XMM workshop in Madrid for hospitality and a warm atmosphere. 
\end{acknowledgements}
%
%
%

%

\begin{thebibliography}{}
\bibitem{} Bardeen, J.~M.: 1970, ApJ~161, 103
\bibitem{} Bardeen, J.~M.: 1970, ApJ~162, 71
\bibitem{} Baumgardt, H., Makino, J., Hut, P., McMillan, S., Portegies Zwart, S.: 2003, ApJ~589, L25
\bibitem{} Bolton, C.~T.: 1972, Natur~235, 271
\bibitem{} Corbin, M.~R.: 1997, ApJ~485, 517
\bibitem{} Cunningham, C.~T.: 1975, ApJ~202, 788
\bibitem{} Fabbiano, G.: 1989, ARA\&A~27, 87
\bibitem{} Fabian, A.~C., Rees, M.~J., Stella, M., White, N.~E.: 1989, MNRAS~238, 729
\bibitem{} Gebhardt, K., Rich, R.~M., Ho, L.~C.: 2005, ApJ~634, 109
\bibitem{} King, A.~R., Davies, M.~B., Ward, M.~J., Fabbiano, G., Elvis, M.: 2001, ApJ~552, L109
\bibitem{} Kollatschny, W.: 2003, A\&A~412, L61
\bibitem{} Laor, A.: 1991, ApJ~376, 90
\bibitem{} Lind, K.~R., Blandford, R.~D.: 1985, ApJ~295, 358
\bibitem{} Lynden--Bell, D.: 1969, Natur~223, 690
\bibitem{} Lynden--Bell, D., Rees, M.~J.: 1971, Natur~152, 461
\bibitem{} Martocchia, A., Matt, G., Karas, V., Belloni, T., Feroci, M.: 2002, A\&A~387, 215
\bibitem{} Miller, J.~M., Fabian, A.~C., Nowak, M.~A., Lewin, W.~H.~G.: 2005, proceedings of the Tenth Marcel Grossmann Meeting, 2003, eds. M. Novello, S. Perez Bergliaffa, R. Ruffini, World Scientific Publishing, 1296
\bibitem{} Miller, J.~M.: 2005, Ap\&SS~300, 227
\bibitem{} M\"uller, A., Camenzind, M.: 2004, A\&A~413, 861
\bibitem{} M\"uller, A., Wold, M.: 2006, A\&A~457, 485
\bibitem{} Nandra, K.: 2006, MNRAS~368, L62
\bibitem{} Netzer, H.: 1977, MNRAS~181, 89
\bibitem{} Netzer, H.: 2003, ApJ~583, L5
\bibitem{} Pakull, M.~W., Grisé, F., Motch, C.: 2006, Populations in high energy sources in galaxies, proceedings IAU symposium No. 230, 2005, eds. E.~J.~A. Meurs \& G. Fabbiano, CUP, 293
\bibitem{} Peterson, B.~M., Meyers, K.~A., Carpriotti, E.~R., Foltz, C.~B., Wilkes, B.~J., Miller, H.~R.: 1985, ApJ~292, 164
\bibitem{} Popovic, L.~C., Vince, I., Atanackovic--Vukmanovic, O., Kubicela, A.: 1995, A\&A~293, 309
\bibitem{} Rees, M.~J.: 1984, ARA\&A~22, 471
\bibitem{} Shapiro, S.~L.: 2005, ApJ~620, 59
\bibitem{} Tanaka, Y., Nandra, K., Fabian, A.~C., et al.: 1995, Natur~375, 659
\bibitem{} Thorne, K.~S.: 1974, ApJ~191, 507
\bibitem{} van der Klis, M.: 1994, ApJS~92, 511
\bibitem{} van der Klis, M.: 2000, ARA\&A~38, 717
\bibitem{} Woltjer, L.: 1959, ApJ~130, 38
\bibitem{} Zheng, W., Sulentic, J.~W.: 1990, ApJ~350, 512
\end{thebibliography}
\end{document}